
%
%
%
\input phyzzx
\tolerance=1000
\sequentialequations
\def\rl{\rightline}

\def\t1{{\tilde 1}}

\def\AEF{A.E. Faraggi}
\def\DVN{D. V. Nanopoulos}

\def\NPB#1#2#3{Nucl. Phys. B{\bf#1} (19#2) #3}
\def\PLB#1#2#3{Phys. Lett. B{\bf#1} (19#2) #3}
\def\PRD#1#2#3{Phys. Rev. D{\bf#1} (19#2) #3}
\def\PRL#1#2#3{Phys. Rev. Lett.{\bf#1} (19#2) #3}

\def\MODA#1#2#3{Mod. Phys. Lett. A{\bf#1} (19#2) #3}
\def\IJMP#1#2#3{Int. J. Mod. Phys. A{\bf#1} (19#2) #3}

\REF\heterotic{D.J. Gross, J.A. Harvey, E. Martinec and
R. Rohm, \PRL{54}{85}{502}; \NPB{256}{85}{253}.}
\REF\CHSW{P.Candelas, G.T. Horowitz, A. strominger and E. Witten,
\NPB{258}{85}{46}.}
\REF\DHVW{L. Dixon, J.A. Harvey, C. Vafa and
E. Witten, \NPB{261}{85}{678}, \NPB{274}{86}{285}.}
\REF\N{Narain, \PLB{169}{86}{41}.}
\REF\FFF{I. Antoniadis, C. Bachas, and C. Kounnas,
\NPB{289}{87}{87}; I. Antoniadis and C. Bachas,
\NPB{298}{88}{586}; H. Kawai, D.C. Lewellen, and S.H.-H. Tye,
\PRL{57}{86}{1832}; \PRD{34}{86}{3794};
\NPB{288}{87}{1}; R. Bluhm, L. Dolan, and P. Goddard,
\NPB{309}{88}{33}.}
\REF\EU{\AEF, \PLB{278}{92}{131} .}
\REF\TOP{\AEF, \PLB{274}{92}{47}.}
\REF\SLM{\AEF, WIS--91/83/NOV--PH; \AEF, WIS--92/16/FEB--PH
(To appear in Nucl.Phys. B.). }
\REF\REVAMP{I. Antoniadis, J. Ellis,
J. Hagelin, and \DVN, \PLB{231}{89}{65}.}
\REF\FNY{\AEF, D.V. Nanopoulos and K. Yuan, \NPB{335}{90}{437}.}
\REF\ALR{I. Antoniadis, G. K. Leontaris and
J. Rizos, \PLB{245}{90}{161}.}
\REF\FOE{\AEF, work in progress.}
\REF\CVETIC{M. Cvetic, \PRL{55}{87}{1795};
\PRL{59}{87}{2829}; \PRL{37}{87}{2366}.}
\REF\KLN{S. Kalara, J. Lopez and D.V. Nanopoulos,
\PLB{245}{91}{421}; \NPB{353}{91}{650}.}
\REF\FN{A.E. Faraggi and D.V. Nanopoulos, \MODA{6}{91}{61}.}
\REF\DSW{M. Dine, N. Seiberg and E. Witten,
\NPB{289}{87}{585};
J.J. Atick, L.J. Dixon and A. Sen, \NPB{292}{87}{109};
S. Cecotti, S. Ferrara and M. Villasante, \IJMP{2}{87}{1839}.}
\REF\naturalness{A.E. Faraggi and D.V. Nanopoulos,
Texas A \& M University preprint CTP--TAMU--78, ACT--15;
\AEF, Ph.D thesis, CTP--TAMU--20/91, ACT--31.}
\REF\SFMM{J. Lopez and \DVN, \NPB{338}{90}{73},
\PLB{251}{90}73; \PLB{256}{91}150; \PLB{268}{91}359;
J. Rizos and K. Tamvakis, \PLB{251}{90}{369}.}
\REF\NRT{\AEF, WIS--92/48/JUN--PH.}
\REF\ART{I. Antoniadis, J. Rizos and K. Tamvakis, \PLB{278}{92}{257}.}

\singlespace
\rl{WIS--92/81/OCT--PH}
\rl{\today}
\rl{T}
\normalspace
\smallskip
\titlestyle{\bf{Generation Mass Hierarchy in Superstring
Derived Models}}
\author{Alon E. Faraggi
{\footnote*{e--mail address: fhalon@weizmann.bitnet}}}
\smallskip
\centerline {Department of Physics, Weizmann Institute of Science}
\centerline {Rehovot 76100, Israel}
\titlestyle{ABSTRACT}

I discuss the problem of generation mass hierarchy in the context
of realistic superstring models which are constructed in the free
fermionic formulation. These models correspond to models which are
compactified on $Z_2\times Z_2$ orbifold. I suggest that the hierarchy among
the generations results from horizontal symmetries, which arise from the
compactification.
In particular, I show that in a class of free fermionic standard--like
models, the suppression of the mass terms for the lightest generation
is a general, and unambiguous,  characteristic of these models.
I show that the mixing between the generations is suppressed due to the
horizontal symmetries.
I conclude that these models may potentially
explain the generation mass hierarchy.

\singlespace
\vskip 0.5cm
\endpage
\normalspace
\pagenumber 1

\centerline{\bf 1. Introduction}

One of the most fundamental problems in high energy physics is
the origin
and hierarchy of the fermion masses.
Why are the three fermion generations, which are universal in their
gauge interactions, separated by orders of magnitude in their masses?
In this respect the Standard Model, and
point field theories in general,
can only be considered as successful attempts
to parameterize the observed mass spectrum.
The Standard Model uses thirteen parameters to parameterize the observed
spectrum. Grand Unified Theories (GUTs) reduce the number of free parameters
and can explain inter family relations between some of the masses.
However, GUTs do not explain the reason for the hierarchy between the
generations nor the smallness of the mixing between different generations.
One has to impose additional horizontal symmetries and specific choices of
Higgs fields to try to explain the family hierarchies. Within the context
of point field theories the problem of generation mass
hierarchy looks rather
arbitrary and some guiding principle is still missing.

Unlike point field theories, superstring theories provide a unique
framework to understand the generations mass hierarchy in terms of
symmetries which are derived in specific superstring models.
The consistency of superstring theory imposes a certain number of degrees
of freedom on the models.
In the closed heterotic string [\heterotic], of the 26 right--moving bosonic
degrees of freedom, 16 are compactified on a flat torus and produce the
observable and hidden gauge groups. Six right--moving bosonic degrees of
freedom,combined with six left--moving degrees of freedom, are compactified
on Calabi--Yau manifold [\CHSW], or on an orbifold [\DHVW].
Alternatively, all the extra degrees of freedom, beyond the four space--time
dimensions, can be taken as bosonic [\N], or fermionic [\FFF],
internal degrees of freedom propagating on the string world--sheet.
The different interpretations are expected to be related. Horizontal
symmetries which arise from the compactification
will be responsible for the generations mass hierarchy.

To illustrate how the compactified space may be responsible for creating
the generation mass hierarchy, I consider models which are constructed in
the free fermionic formulation of the heterotic string. Of these,
I focus mainly on a specific class of standard--like models [\EU,\TOP,\SLM].
The standard--like models have several unique characteristics. First,
they have three and only three generations of chiral fermions. The
chiral generation states are obtained from the three distinct twisted
sectors of the corresponding orbifold model and none from the untwisted
sector. Second, the standard--like models suggest an explanation for the
heaviness of the top quark relative to the lighter quarks and leptons.
At the trilinear level of the superpotential only the top quark
gets a non vanishing mass term. The bottom quark and the lighter quarks and
leptons get their mass terms from nonrenormalizable terms, which are
suppressed relative to the leading cubic level terms.
Finally, the standard--like models naturally evade the problem with
proton decay from dimension four operators that usually exist in superstring
models which are based on an intermediate GUT symmetry [\SLM].

\bigskip
\centerline{\bf 2. Realistic free fermionic models}

In the free fermionic formulation of the heterotic string
in four dimensions all the world--sheet
degrees of freedom  required to cancel
the conformal anomaly are represented  in terms of free fermions
propagating on the string world--sheet.
Under parallel transport around a noncontractible loop the
fermionic states pick up a phase. A model in this construction
is defined by a set of  basis vectors of boundary conditions for all
world--sheet fermions. These basis vectors are constrained by the string
consistency requirements (e.g. modular invariance) and
completely determine the vacuum structure of
the model. The physical spectrum is obtained by applying the generalized
GSO projections. The low energy effective field theory is obtained
by S--matrix elements between external states. The Yukawa couplings
and higher order nonrenormalizable terms in the superpotential
are obtained by calculating corralators between vertex operators.
For a corralator to be nonvanishing all the symmetries of the model must
be conserved. Thus, the boundary condition vectors determine the
phenomenology of the models.

The first five vectors (including the vector {\bf 1}) in the basis are
$$\eqalignno{S&=({\underbrace{1,\cdots,1}_{{\psi^\mu},
{\chi^{1,...,6}}}},0,\cdots,0
\vert 0,\cdots,0).&(1a)\cr
b_1&=({\underbrace{1,\cdots\cdots\cdots,1}_
{{\psi^\mu},{\chi^{12}},y^{3,...,6},{\bar y}^{3,...,6}}},0,\cdots,0\vert
{\underbrace{1,\cdots,1}_{{\bar\psi}^{1,...,5},
{\bar\eta}^1}},0,\cdots,0).&(1b)\cr
b_2&=({\underbrace{1,\cdots\cdots\cdots\cdots\cdots,1}_
{{\psi^\mu},{\chi^{34}},{y^{1,2}},
{\omega^{5,6}},{{\bar y}^{1,2}}{{\bar\omega}^{5,6}}}}
,0,\cdots,0\vert
{\underbrace{1,\cdots,1}_{{{\bar\psi}^{1,...,5}},{\bar\eta}^2}}
,0,\cdots,0).&(1c)\cr
b_3&=({\underbrace{1,\cdots\cdots\cdots\cdots\cdots,1}_
{{\psi^\mu},{\chi^{56}},{\omega^{1,\cdots,4}},
{{\bar\omega}^{1,\cdots,4}}}},0,\cdots,0
\vert {\underbrace{1,\cdots,1}_{{\bar\psi}^{1,...,5},
{\bar\eta}^3}},0,\cdots,0).&(1d)\cr}$$
with the choice of generalized GSO projections
$$c\left(\matrix{b_i\cr
                                    b_j\cr}\right)=
c\left(\matrix{b_i\cr
                                    S\cr}\right)=
c\left(\matrix{1\cr
                                    1\cr}\right)=-1,\eqno(2)$$
and the others given by modular invariance.
This set is reffered to as the NAHE{\footnote*{This set was first
constructed by Nanopoulos, Antoniadis, Hagelin and Ellis  (NAHE)
in the construction
of  the flipped $SU(5)$ [\REVAMP].  {\it nahe}=pretty, in
Hebrew.}} set.
The NAHE set is common to all  the realistic models constructed  in the
free fermionic formulation [\REVAMP,\FNY,\ALR,\EU,\TOP,\SLM,\naturalness]
and is a basic set common to all the models which I discuss.
The sector {\bf S} generates  $N=4$ space--time supersymmetry, which is broken
to $N=2$ and $N=1$ space--time supersymmetry by $b_1$ and $b_2$, respectively.
Restricting $b_j\cdot S=0$mod$2$, and $c\left(\matrix{S\cr
                                                b_j\cr}\right)=\delta_{b_j}$,
for all basis vector $b_j\epsilon{B}$ guarantees the existence of
$N=1$ space--time supersymmetry. The superpartners from a given sector
$\alpha\epsilon\Xi$ are obtained from the sector $S+\alpha$.
The gauge group after the NAHE set is $SO(10)\times
E_8\times SO(6)^3$ with $N=1$ space--time supersymmetry.
The three $SO(6)$
symmetries are horizontal, generational dependent, symmetries.

Models based on the NAHE set correspond to models that are based on
$Z_2\times Z_2$ orbifold.
This correspondence
is illustrated by extending the  $SO(10)$
symmetry to $E_6$. Adding the vector
$$X=(0,\cdots,0\vert{\underbrace{1,\cdots,1}_{{{\bar\psi}^{1,\cdots,5}},
{{\bar\eta}^{1,2,3}}}},0,\cdots,0)\eqno(3)$$
to the NAHE set, extends the gauge symmetry to
$E_6\times U(1)^2\times SO(4)^3$.
The set $\{{\bf 1},{\bf S},I={\bf 1}+b_1+b_2+b_3,X\}$ produces a
$E_8\times E_8$ toroidal compactification on a $SO(12)$ lattice.
The $SO(12)$ symmetry is reproduced for special values of the metric and
the antisymmetric tensor. The metric, $g_{ij}$, is given by the cartan
matrix of $SO(12)$ and the antisymmetric tensor, $b_{ij}$, is given by
$$b_{ij}=\cases{
g_{ij}&;\ $i>j$,\cr
0&;\ $i=j$,\cr
-g_{ij}&;\ $i<j$.\cr}\eqno(4)$$
The sectors $b_1$ and $b_2$ correspond to the $Z_2\times{Z_2}$ twist and break
the symmetry to $SO(4)^3\times{E_6}\times{U(1)^2}\times{E_8}$.
The fermionic states $\{\chi^{12},\chi^{34},\chi^{56}\}$ and
$\{{\bar\eta}^1,{\bar\eta}^2,{\bar\eta}^3\}$ give the usual ``standard--
embedding", with
$b(\chi^{12},\chi^{34},\chi^{56})=b({\bar\eta}^1,{\bar\eta}^2,{\bar\eta}^3)$.
The $U(1)$ current of the left--moving $N=2$ world--sheet supersymmetry
is given by $J(z)=i\partial_z(\chi^{12}+\chi^{34}+\chi^{56})$,
and the $U(1)$ charges in the decomposition of $E_6$ under
$SO(10)\times U(1)$ are given by the world--sheet current
${\bar\eta}^1{\bar\eta}^{1^*}+{\bar\eta}^2{\bar\eta}^{2^*}
+{\bar\eta}^3{\bar\eta}^{3^*}$.
The sectors $(b_1;b_1+X)$, $(b_2;b_2+X)$ and $(b_3;b_3+X)$
each give eight $27$ of $E_6$, and correspond to the twisted sectors
of the orbifold model. The $(NS;NS+X)$ sector gives in
addition to the vector bosons and spin two states, three copies of
scalar representations in $27+{\bar {27}}$ of $E_6$. This sector corresponds
to the untwisted sector of the orbifold model.

In this model the only internal fermionic states which count the
multiplets of $E_6$ are the real internal fermions $\{y,w\vert{\bar y},
{\bar\omega}\}$. This is observed by writing the degenerate vacuum
of the sectors $b_j$ in a combinatorial notation. The vacuum of the sectors
$b_j$  contains twelve periodic fermions. Each periodic fermion
gives rise to a two dimensional degenerate vacuum $\vert{+}\rangle$ and
$\vert{-}\rangle$ with fermion numbers $0$ and $-1$, respectively.
The GSO operator, is a generalized parity operator, which
selects states with definite parity. After applying the
GSO projections, we can write the degenerate vacuum of the sector
$b_1$ in combinatorial form
$$\eqalignno{\left[\left(\matrix{4\cr
                                    0\cr}\right)+
\left(\matrix{4\cr
                                    2\cr}\right)+
\left(\matrix{4\cr
                                    4\cr}\right)\right]
\left\{\left(\matrix{2\cr
                                    0\cr}\right)\right.
&\left[\left(\matrix{5\cr
                                    0\cr}\right)+
\left(\matrix{5\cr
                                    2\cr}\right)+
\left(\matrix{5\cr
                                    4\cr}\right)\right]
\left(\matrix{1\cr
                                    0\cr}\right)\cr
+\left(\matrix{2\cr
                                    2\cr}\right)
&\left[\left(\matrix{5\cr
                                    1\cr}\right)+
\left(\matrix{5\cr
                                    3\cr}\right)+
\left(\matrix{5\cr
                                    5\cr}\right)\right]\left.
\left(\matrix{1\cr
                                    1\cr}\right)\right\}&(5)\cr}$$
where
$4=\{y^3y^4,y^5y^6,{\bar y}^3{\bar y}^4,
{\bar y}^5{\bar y}^6\}$, $2=\{\psi^\mu,\chi^{12}\}$,
$5=\{{\bar\psi}^{1,\cdots,5}\}$ and $1=\{{\bar\eta}^1\}$.
The combinatorial factor counts the number of $\vert{-}\rangle$ in the
degenerate vacuum of a given state.
The two terms in the curly brackets correspond to the two
components of a Weyl spinor.  The $10+1$ in the $27$ of $E_6$ are
obtained from the sector $b_j+X$.
{}From Eq. (5) it is observed that the states
which count the multiplicities of $E_6$ are the internal
fermionic states $\{y^{3,\cdots,6}\vert{\bar y}^{3,\cdots,6}\}$.
A similar result is
obtained for the sectors $b_2$ and $b_3$ with $\{y^{1,2},\omega^{5,6}
\vert{\bar y}^{1,2},{\bar\omega}^{5,6}\}$
and $\{\omega^{1,\cdots,4}\vert{\bar\omega}^{1,\cdots,4}\}$
respectively, which suggests that
these twelve states correspond to a six dimensional
compactified orbifold with Euler characteristic equal to 48.
The number of fixed points in the $Z_2\times{Z_2}$ orbifold on a $SO(12)$
lattice is 48 and matches twice the number of generations in the fermionic
model. The correspondence between the fermionic models and the
$Z_2\times{Z_2}$ orbifold will be discussed further in Ref. [\FOE].
The important point to realize is that in the fermionic formulation
the 12 internal fermionic states, $\{y,w\vert{\bar y},
{\bar\omega}\}$, play the role of the six dimensional ``compactified space''
of the orbifold. The boundary conditions, assigned to these internal fermions,
determine many of the properties of the low energy spectrum.

Turning back to the NAHE set. At the level of the NAHE set,
the sectors $b_1$,  $b_2$ and $b_3$ each
produce 16 chiral generations. There exists a permutation symmetry between
these sectors. The $SO(6)^3$  horizontal symmetries constrain the
possible interactions.

The number of generations is reduced by adding three additional vectors
to the NAHE set. The standard--like models use $Z_2\times{Z_2{\times{Z_4}}}$,
where the $Z_4$ twist is used to break
$SO(2n)\rightarrow SU(n)\times U(1)$. The boundary
conditions of the gauge sector are fixed by requiring that the $SO(10)$
symmetry breaks to the Standard Model,
and by modular invariance constraints [\SLM].
The assignment of boundary conditions to the set of real internal fermions
$\{y,\omega\vert{\bar y},{\bar\omega}\}$ distinguishes between different
models. The possible assignments are constrained by requiring a net chirality
of three generations. One half of the generations is projected out
by the $Z_4$
twist. The assignment of boundary conditions to the real fermions,
$\{y,\omega\vert{\bar y},{\bar\omega}\}$, is constrained by requiring that
the combinatorial factor in Eq. (5) reduces to one, for each of the $b_1$,
$b_2$ and $b_3$ sectors. At the same time the three $SO(6)$ horizontal
symmetries are broken to horizontal $U(1)$ symmetries. Three U(1)s,
$U(1)_{r_j}$ $(j=1,2,3)$,
correspond to the world--sheet currents ${\bar\eta}_1{\bar\eta}_1^*$,
${\bar\eta}_2{\bar\eta}_2^*$ and ${\bar\eta}_3{\bar\eta}_3^*$. These
$U(1)$ symmetries are a generic feature of realistic free fermionic
models [\REVAMP,\FNY,\ALR,\EU,\TOP,\SLM].
Additional $U(1)$ symmetries arise from complexification
of real right--moving fermions from the set $\{{\bar y},{\bar\omega}\}$.
In the standard--like models, requiring that the Higgs doublets
from the Neveu--Schwarz sector survive the GSO projections,
imposes at least three additional $U(1)$ symmetries [\SLM]. One for each
sector $b_1$, $b_2$ and $b_3$. In the models of tables 1 (model 1) [\EU]
and 2 (model 2) [\TOP], they
correspond to the right--moving world--sheet currents
${\bar y}_3{\bar y}_6$, ${\bar y}_1{\bar\omega}_5$ and
${\bar\omega}_2{\bar\omega}_4$, denoted by $U(1)_{r_{j+3}}$
$(j=1,2,3)$. Other choices for the low energy gauge
group do not impose such a restriction. In addition to these symmetries,
for every right--moving $U(1)$ symmetry correspond
a left--moving global $U(1)$ symmetry. The first three,
$U(1)_{\ell_j}$
$(j=1,2,3)$,
correspond to the charges of the supersymmetry generator
$\chi^{12}$, $\chi^{34}$ and $\chi^{56}$. These are common to all the
realistic free fermionic models. Additional global $U(1)$ symmetries
arise from additional complexified left--moving fermions. In the
standard--like models of tables 1 and 2,
the last three,
$U(1)_{\ell_{j+3}}$ $(j=1,2,3)$,
correspond to the complexified left--moving fermions
$y^3y^6$, $y^1\omega^5$ and $\omega^2\omega^4$.
Finally, the models contain Ising model sigma operators,
which are obtained by pairing a left--moving
real fermion with a right--moving real fermion. In the standard--like models
[\EU,\TOP] there are six Ising model operators,
$\sigma^i_\pm=\{\omega^1{\bar\omega}^1,
y^2{\bar y}^2, \omega^3{\bar\omega}^3, y^4{\bar y}^4,
y^5{\bar y}^5, \omega^6{\bar\omega}^6\}_\pm$. These symmetries are additional
horizontal symmetries, which constrain the possible  F--terms
in the superpotential. Thus, each sector
$b_1$, $b_2$ and $b_3$ produces one generation, with horizontal symmetries.
The notation used in tables 1 and 2, emphasizes the division of the internal
world--sheet fermions among the three generations.

Trilinear and
nonrenormalizable contributions to the superpotential are obtained
by calculating corralators between vertex operators [\CVETIC,\KLN],
$$A_N\sim\langle V_1^fV_2^fV_3^b\cdot\cdot\cdot V_N^b\rangle,\eqno(6)$$
where $V_i^f$ $(V_i^b)$ are the fermionic (scalar)
components of the vertex operators.
The non vanishing terms are obtained by
applying the rules of Ref. [\KLN].
To obtain the correct ghost charge some of the
vertex operators are picture
changed by taking
$$V_{q+1}(z)=\lim_{w\to z}exp(c)(w)T_F(w)V_{q}(z),\eqno(7)$$
where $T_F$ is the super current and in the
fermionic construction is given by
$$T_F=\psi^\mu\partial_\mu X+i{\sum_{I=1}^6}\chi_{_I}{y_{_I}}
\omega_{_I}=T_F^0+T_F^{-1}+T_F^{+1}\eqno(8)$$
with
$$T_F^{-1}=e^{-i\chi^{12}}\tau_{_{12}}+e^{-i\chi^{34}}\tau_{_{34}}
+e^{-i\chi^{56}}\tau_{_{56}}{\hskip .5cm};
{\hskip .5cm}T_F^{-1}=(T_F^{+1})^*\eqno(9)$$
where
$\tau_{_{ij}}={i\over{\sqrt2}}(y^i\omega^i+y^j\omega^j)$
and $e^{\chi^{ij}}={1\over\sqrt{2}}(\chi^i+i\chi^j)$.

Several observations simplify the analysis of the
potential non vanishing
terms. First, it is observed that only the $T_F^{+1}$ piece of $T_F$
contributes to $A_N$ [\KLN]. Second, in the standard--like models
[\EU,\TOP] the pairing of left--moving fermions is
$y^1\omega^5$, $\omega^2\omega^4$ and $y^3y^6$.
One of the fermionic states in every term $y^i\omega^i$ $(i=1,...,6)$
is
complexified and therefore can be written,
for example for $y^3$ and $y^6$,
as
$$y^3={1\over{\sqrt2}}(e^{iy^3y^6}+e^{-iy^3y^6}),
y^6={1\over{\sqrt2}}(e^{iy^3y^6}-e^{-iy^3y^6}).\eqno(10)$$
Consequently, every picture changing operation changes the
total
$U(1)_\ell=U(1)_{\ell_4}+U(1)_{\ell_5}+U(1)_{\ell_6}$ charge
by $\pm1$. An odd (even)  order term
requires an even (odd) number of picture changing
operations to get the correct ghost number [\KLN].
Thus, for $A_N$ to be non vanishing,
the total $U(1)_\ell$ charge, before picture
changing, has to be an odd (even)
number, for even (odd) order terms, respectively.
Similarly, in every pair $y_i\omega_i$, one real
fermion, either $y_i$ or
$\omega_i$, remains real and is paired with the corresponding
right--moving real fermion to produce an Ising model sigma operator.
Every picture changing operation changes the number of left--moving
real fermions by one.
This property of the standard--like models significantly
reduces the number of
potential non vanishing terms.

\bigskip
\centerline{\bf 3.  Higgs mass matrix }

There are two types of Higgs doublets, from two distinct sectors,
common to all the realistic free fermionic models
[\REVAMP,\ALR,\EU,\TOP,\SLM].
The first type are Higgs doublets form the Neveu--Schwarz sector.
They correspond to Higgs doublets from the untwisted sector in the
orbifold language.
In the standard--like models [\FNY,\EU,\TOP,\SLM] and the
$SO(6)\times SO(4)$ [\ALR] models, the presence of Higgs
doublets from the untwisted sector in the massless spectrum, is correlated
with the additional $U(1)_{r_{j+3}}$ $(j=1,2,3)$ horizontal
symmetries, which arise from pairing real right--moving fermions [\SLM].
In all the realistic standard--like models that are based on the NAHE
set, there are three pairs of Higgs doublets $h_1$, ${\bar h}_1$
$h_2$, ${\bar h}_2$ and  $h_3$, ${\bar h}_3$, from the untwisted sector.
Each pair $h_j$, ${\bar h}_j$ carries $U(1)_{r_j}$ charge and therefore
can couple
at the cubic level only to the states from the sector $b_j$.

The second type of Higgs doublets is obtained from a combination of the
two $Z_2\times Z_2$ vectors,
which are used to reduce the number of generations,
and some combination of $b_1$, $b_2$ and $b_3$. In the flipped $SU(5)$
 [\REVAMP]  and the $SO(6)\times SO(4)$ model [\ALR] the
combination is $b_4+b_5$. In the standard--like models
 of tables 1 and 2, the combination is
$\zeta=b_1+b_2+\alpha+\beta$. In this vector,
$\zeta_R\cdot{\zeta_R}=\zeta_L\cdot{\zeta_L}=4$.
 Therefore, the massless states are obtained by acting on the vacuum
 with one right--moving fermionic oscillator. The states in this sector
 transform only under the observable gauge group. The presence of these
states in the massless spectrum, and consequently of the vector
combination in the additive group is essential for the application of the
Dine--Seiberg--Witten (DSW) mechanism [\DSW] and for obtaining realistic
phenomenology. Requiring the
existence of this vector combination in the additive group is an additional
strong constraint on the allowed basis vectors, which extend the NAHE set.
These two $Z_2\times Z_2$ basis vectors play an important role
in generating the  generation mass hierarchy. Their combination is
symmetric with respect to $b_1$ and $b_2$. However, they brake the
cyclic symmetry between $b_1$, $b_2$ and $b_3$.

The light Higgs spectrum is determined by the massless eigenstates of the
doublet Higgs mass matrix. The doublet mass matrix consists of the terms
$h_i{\bar h}_j\langle\Phi^n\rangle$, and is defined by
$h_i(M_h)_{ij}{\bar h}_j$, $i,j=1,2,3,4$
where $h_i=(h_1,h_2,h_3,h_{45})$
and ${\bar h}_i=({\bar h}_1,{\bar h}_2,{\bar h}_3,{\bar h}_{45})$.
At the cubic level of the
superpotential the Higgs doublets mass matrix is given by,
$$M_h={\left(\matrix{0&{\bar\Phi}_{12}&{{\bar\Phi}_{13}}&0\cr
                     \Phi_{12}&0&{{\bar\Phi}_{23}}&0\cr
                     {\Phi_{13}}&\Phi_{23}&0&{\Phi_{45}}\cr
                      0&0&{\bar\Phi}_{45}&0\cr}\right).}\eqno(11)$$

At the cubic level this form of the Higgs mass matrix is common to
the flipped $SU(5)$ string model [\REVAMP] and to the
realistic standard--like models, which are based on the NAHE set
[\EU,\TOP,\SLM].
These models contain an anomalous $U(1)$ symmetry, which breaks
supersymmetry at the Planck scale,
and destabelizes the vacuum. Supersymmetry is restored
by giving a VEV to some singlets in  the spectrum, along F and D flat
directions [\DSW]. In the flipped $SU(5)$ string model and the
standard--like models, it has been found
that we must impose [\SFMM,\EU,\TOP,\NRT],
$$\langle{\Phi_{12},{\bar\Phi}_{12}}\rangle=0,\eqno(12)$$
and that $\Phi_{45}$, and ${\bar\Phi}_{13}$ or ${\bar\Phi}_{23}$,
must be different from zero. From this result it follows that in any
flat F and D solution, $h_3$ and ${\bar h}_3$ obtain a Planck scale mass.
This result is not surprising. It is a consequence of the symmetry of the
sectors $\alpha$ and $\beta$ with respect to the $b_1$ and $b_2$ sectors.
The implication is that $h_3$ and ${\bar h}_3$ do not contribute to
the light Higgs representations. Consequently, the mass terms for the
states from the sector $b_3$ will be suppressed.

The matrix $M_h$ is diagonalized by $S{M_h}T^{\dagger}$ where $S$ and $T$
are two unitary matrices and $(SM_hT^\dagger)_{ij}=m_i\delta_{ij}$.
It follows that $SM{M^\dagger}S=TM{^\dagger}MT=\vert{m}\vert^2$.
The $h$ and ${\bar h}$
mass eigenstates are obtained by evaluating the
eigenvalues and eigenstates
of $MM{^\dagger}$ and $M{^\dagger}M$, respectively.
At the cubic level of the superpotential there are two pairs of light Higgs
states. Additional vanishing terms in the cubic level Higgs mass matrix depend
on specific F and D flat solutions. At the nonrenormalizable level of the
superpotential, additional non vanishing entries in the Higgs mass matrix
can appear. For example in the standard--like model of table 1,
at the quintic level we get,
$$h_2{\bar h}_{45}\Phi_{45}H_{25}H_{26}{\hskip 1cm};{\hskip 1cm}
  {\bar h}_2h_{45}{\bar\Phi}_{45}H_{23}H_{27}.\eqno(13a,b)$$
These additional terms reduce the number of light Higgs pairs
to one pair. For example, if
$\langle{H_{25}}\rangle\sim
\langle{H_{26}}\rangle\sim{10^{14}}GeV$, one of the light pairs
receives a mass of $O(10^{10}GeV)$. The light eigenstates are
${\bar h}_2$ and $h_{45}$. A VEV for ${\bar\Phi}_{45}$ of the
order of ${\langle{\bar\Phi}_{45}\rangle}\sim O(10^{10}GeV)$, produces
the mixing between the two light Higgs eigenstates.
At order $N=7$ we obtain
additional terms of the form $h_1{\bar h}_2V_iV_j\phi^3$, where $V_iV_j$
is a condensate of the hidden $SU(5)$ gauge group.
 These terms make the extra pair massive without
breaking $U(1)_{Z^\prime}$. They are proportional to
$({{\Lambda_5}\over{M}})^2$, where $\Lambda_5$ is the scale at which the
hidden $SU(5)$ group is strongly interacting.
The remaining light combinations depend on the specific
entries in the Higgs mass matrix which become non zero and depend on
specific F and D flat solutions. However, already at this stage,
and without knowledge of the specific solution, we can see how the
symmetries of the spin structure are reflected in the generation
mass hierarchy.

Among the realistic free fermionic models, the standard--like models
[\FNY,\EU,\TOP,\SLM]
have the unique property that the generations from the
twisted sectors $b_1$,
$b_2$ and $b_3$ are the only light generations. There aren't additional
generations and mirror generations, which become massive at some high scale.
Thus, the identification of the light generations is unambiguous.
Below, I focus entirely on this class of standard--like models.

\bigskip
\centerline{\bf 4.  Generation mass hierarchy}

The class of superstring
standard--like models have two unique properties that restrict the
fermion mass terms. A unique property
of the standard--like models is the possible connection between the
requirement of a supersymmetric vacuum at the Planck scale, via the
DSW mechanism, and the heaviness of the top quark relative to the
lighter quarks and leptons [\SLM]. The only standard--like models
which were found to  admit a solution to the set of F and D constraints
are models in which only $+{2\over3}$ charged quarks obtain trilevel Yukawa
couplings [\SLM]. Trilevel Yukawa couplings for $+{2\over3}$ or
$-{1\over3}$ charged quarks are selected by the assignment of boundary
conditions for the real fermions in the vector $\gamma$. They are
determined by [\SLM],
$$\Delta_j=\vert\gamma(U(1)_{\ell_{j+3}})-\gamma(U(1)_{r_{j+3}})\vert=0,1
{\hskip 1.2cm}(j=1,2,3).\eqno(14)$$
$\Delta_j=0$ gives a trilevel Yukawa coupling for $-{1\over3}$ charged
quarks and $\Delta_j=1$ gives a Yukawa coupling for $+{2\over3}$ charged
quarks. The only standard--like models that admit F and D flat solution
are models with $\Delta_j=1$ for the three sectors $b_1$, $b_2$ and
$b_3$. The second property unique to the standard--like models is
the unambiguous identification of the light generations. There are
only three twisted generations from the sectors $b_1$, $b_2$ and $b_3$
and none from the untwisted sector.

The symmetry of the vectors $\alpha$ and $\beta$ with respect to the
vectors $b_1$ and $b_2$, forces $h_3$ and ${\bar h}_3$ to get a Planck scale
mass. Nonrenormalizable terms have the form
$cg^{N-2}f_if_jh\phi^{^{N-3}}(2\alpha^\prime)^{N-3}$, or
 $cg^{N-2}f_if_j{\bar h}\phi^{^{N-3}}(2\alpha^\prime)^{N-3}$,
where $f_i$, $f_j$ are two fermions from
the sectors $b_1$, $b_2$ and $b_3$. $h$ and ${\bar h}$ are Higgs doublets
which are combinations of $(h_1,h_2,h_{45})$ and
$({\bar h}_1,{\bar h}_2,{\bar h}_{45})$, respectively.
The coefficients, $c=O(1)$, can be obtained by calculating the nontrivial
corralators between the vertex operators, and $g$ is the gauge coupling
constant. The combination
$\phi^{^{N-3}}$ is a combination of fields that get a VEV. Using the relation
${1\over2}g\sqrt{\alpha^\prime}={{\sqrt{8\pi}}/{M_{Pl}}}$, the
nonrenormalizable terms take the form,
$cgf_if_jh{({{\langle\phi\rangle}\over{M}})^{^{N-3}}}$.
Thus, the
nonrenormalizable terms become effective trilinear terms, suppressed by
${({{\langle\phi\rangle}\over{M}})^{^{N-3}}}$ relative to the trilevel terms,
where $M\equiv{{M_{Pl}}\over{2\sqrt{8\pi}}}\sim{10^{18}GeV}$ [\KLN].
In the standard--like models several scales contribute to these
generalized VEVs: (a) The leading scale is the scale of singlet VEVs,
${{\langle\phi\rangle}\over{M}}$, which
are used to cancel the D--term equation of the anomalous $U(1)_A$.
These are typically of the order of
${{\langle\phi\rangle}\over{M}}\sim{1\over{10}}$.
(b) The scale of hidden sector condensates, ${{\langle{TT}\rangle}\over
{M^2}}$. The hidden sector
contains two non abelian gauge groups, $SU(5)\times SU(3)$, with
$\Lambda_5>>\Lambda_3$, and $\Lambda_5\ge10^{14}GeV$. Thus, the
leading terms are proportional to hidden $SU(5)$ condensates, and the
analysis focuses on these terms.
(c) The scale of $Z^\prime$ breaking. In Ref. [\NRT], it was shown
that VEVs that break $U(1)_{Z^\prime}$ violate the cubic level F flat
solution and therefore break supersymmetry at the Planck scale. Therefore,
$\Lambda_{Z^\prime}\le\Lambda_5$.

I now turn to examine the fermion mass terms in the standard--like models.
At the cubic level the only potential terms are $u_1Q_1{\bar h}_1$
and $u_2Q_2{\bar h}_2$ [\EU,\TOP].
Below the intermediate scale ${\bar h}_1$
or ${\bar h}_2$ obtain a large mass and one term remains. Thus,
only the top quark has a cubic level mass term, and only its mass is
characterized by the electroweak scale. This property is common to
all the standard--like models which admit supersymmetric, F and D flat,
solutions at the Planck scale.

The quartic and quintic orders mass terms differ between the models
of tables 1 and 2. This again is a consequence of the assignment
of boundary conditions in the vectors $\alpha$ and $\beta$ [\SLM].
The boundary conditions in the gauge sector,
$\{{\bar\psi}^{1,\cdots,5},{\bar\eta}^{1,2,3},{\bar\phi}^{1,\cdots,8}\}$,
are identical in the two
models. The two models differ by the boundary conditions
of the internal fermions $\{y,w\vert{\bar y},{\bar\omega}\}$. This
is reflected in the nonvanishing higher order mass terms.

In model 1 there are no potential quark and lepton
mass terms at the quartic order.
At the quintic order we get the nonvanishing terms,
$$\eqalignno{&d_2Q_2h_{45}{\bar\Phi}_2^-\xi_1,{\hskip .2cm}
       e_2L_2h_{45}{\bar\Phi}_2^+\xi_1&(15a)\cr
      &d_1Q_1h_{45}{\Phi}_1^+\xi_2,{\hskip .2cm}
       e_1L_1h_{45}{\Phi}_1^-\xi_2&(15b)\cr
      &u_2Q_2({\bar h}_{45}\Phi_{45}{\bar\Phi}_{23}+
  {\bar h}_1{\bar\Phi}_i^+{\bar\Phi}_i^-)&(15c)\cr
      &u_1Q_1({\bar h}_{45}\Phi_{45}{\bar\Phi}_{13}+
  {\bar h}_2{\Phi}_i^+{\Phi}_i^-)&(15d)\cr
      &(u_2Q_2h_2+u_1Q_1h_1)
  {{\partial W}\over{\partial\xi}_3}.&(15e)\cr}$$
These terms can produce mass terms for the charm quark and for the two
heavier generations of $-{1\over3}$ charged quarks and for charged
leptons.

In model 2 we get at the quartic order potential mass terms for
$-{1\over3}$ charged quarks and for charged leptons, from the sectors
$b_1$ and $b_2$,
$${d_{L_1}^c}Q_1h_{45}^\prime\Phi_1,{\hskip .2cm}
{e_{L_1}^c}L_1h_{45}^\prime\Phi_1,{\hskip .2cm}
{d_{L_2}^c}Q_2h_{45}^\prime{\bar\Phi}_2,{\hskip .2cm}
{e_{L_2}^c}L_2h_{45}^\prime{\bar\Phi}_2,\eqno(16)$$
while there are no non vanishing quartic order terms for $+{2\over3}$
charged quarks. At order $N=5$ potential mass terms appear for the
charm quark of the form
$u_2Q_2({\bar h}_{45}\phi^2+{\bar h}_1\phi^2+{\bar h^{\prime}}_{45}\phi^2)$
where $\phi^2$ represent combinations of singlet VEVs. There are no potential
quintic order mass terms for $-{1\over3}$ charged quarks and for charged
leptons.
At order $N=6$, there are additional terms of the form $e_iL_ih\phi^3$ and
$d_iQ_ih\phi^3$ (i=1,2), which produce possible diagonal mass terms for the
strange quark and for the $\mu$ lepton.

At this stage it is seen that the mass terms for the $b_1$ and $b_2$
sectors come from terms which are suppressed by powers of
$({{\langle\phi\rangle}\over{M}})^{N-3}$, where $\phi$ are singlets VEVs
that are used to cancel the anomalous D--term equation.
The split between these two sectors in terms
of the boundary condition vectors is still
not transparent. As discussed above the vectors
$\alpha$ and $\beta$ are symmetric with respect to $b_1$ and $b_2$.
Thus, the symmetry can be broken by the vector $\gamma$. Examination
of tables 1 and 2 reveals that this is the case in model 2, while
in model 1 the vector $\gamma$ is symmetric with respect to $b_1$ and
$b_2$. Thus, in this model the symmetry between the sectors $b_1$
and $b_2$ has to be broken by the choice of generalized GSO
phases or by specific choices of flat directions.

Rather than the presence of potential leading mass terms for $G_1$
and $G_2$, the most important aspect of nonrenormalizable terms
is the absence of such terms for $G_3$. As argued above the Higgs
doublets $h_3$ and ${\bar h}_3$ get a Planck scale mass and do not
contribute to the light Higgs representations. Similarly, requiring
F and D flat solution to the anomalous D--term equation imposes
[\EU,\TOP,\NRT],
$$\langle\Phi_{12},{\bar\Phi}_{12},\xi_3\rangle\equiv0.\eqno(17)$$
The potential leading terms for $G_3$ have the form $f_3f_3h\phi^{^{N-3}}$
or $f_3f_3{\bar h}\phi^{^{N-3}}$, where $f_3$ are fermions from the
sectors $b_3$, $h$ and ${\bar h}$ are combinations of
$\{h_1,h_2,h_{45}\}$ and $\{{\bar h}_1,{\bar h}_2,{\bar h}_{45}\}$
respectively, and $\phi^{^{N-3}}$ is a combination of singlets VEVs.
However, each $f_3$ carries $U(1)_{\ell_3}={1\over2}$ to give a total
of $U(1)_{\ell_3}=1$. The only singlets that do not break $U(1)_{Z^\prime}$
and which have $U(1)_{\ell_3}$ charge are $\Phi_{12}$, ${\bar\Phi}_{12}$
and $\xi_3$. Consequently, all the potential leading mass terms for $G_3$
vanish identically to all orders. A possible way to get  diagonal mass
terms for $G_3$ is to couple them with the states from the
$b_j+2\gamma$ sectors, which generate $SU(5)$ condensates, or
with VEVs that break $U(1)_{Z^\prime}$.
For example, in model 1 these states come from the sectors
$b_1+b_3+\alpha\pm\gamma+(I)$ and $b_2+b_3+\beta\pm\gamma+(I)$.
In this model at the quintic level we get a term
$u_3Q_3{\bar h}_{45}H_{17}H_{24}$.
However, as argued above $\Lambda_{Z^{\prime}}\le\Lambda_5$. Therefore,
the diagonal mass terms for $G_3$ are suppressed by at least
$({{\Lambda_5}\over{M}})^2\lambda_t$, where $\lambda_t$
is the top Yukawa coupling.
Thus, the states from $G_3$ are identified with the lightest
generation. The suppression of their mass terms is a consequence
of the symmetries of the vectors which extend the NAHE set.
I would like to emphasize that this is a general result which will be
applicable to all the realistic standard--like models  [\EU,\TOP,\SLM],
and may be a general result of realistic free fermionic models.
A similar result holds in the flipped $SU(5)$ string model [\SFMM,\ART].
However, there, one needs to avoid identifying $G_3$ with the lightest
generation
because of problems with dimension four operators, which mediate rapid
proton decay [\ART].

Next I turn to examine the mixing between the generations. The mixing terms
have the form $f_if_jh\phi^n$ and $f_if_j{\bar h}\phi^n$, where
$i\ne j$,  $h$ and ${\bar h}$ are light Higgs combinations and $\phi^n$
is a combination of generalized VEVs.

The fermion states from each sector $b_j$ carry
$U(1)_{\ell_{j+3}}=\pm{1\over2}$.
The singlets from the NS sector and the sector $b_1+b_2+\alpha+\beta$,
all have $U(1)_{\ell_{j+3}}=0$. Every picture changing operation changes
the total $U(1)_\ell=U(1)_{\ell_4}+U(1)_{\ell_5}+U(1)_{\ell_6}$
by $\pm1$. Thus, to construct nonrenormalizable terms which are invariant
under $U(1)_\ell$, we must tag to $f_if_jh$ additional fields
with $U(1)_{\ell_{j+3}}=\pm{1\over2}$. For example, examining model 1 [\EU],
we observe that the only available states are from the sectors
$b_j+2\gamma$. These states transform under the hidden $SU(5)\times SU(3)$
gauge group in the fundamental representations, $5$ and ${\bar 5}$ of $SU(5)$,
 and $3$ and ${\bar 3}$ of $SU(3)$.
Thus, the mixing terms are suppressed by at least
$({{\Lambda_5}\over{M}})^2$ relative to the leading diagonal terms.

\bigskip
\centerline{\bf 5.  Conclusions }

In this paper I examined the texture of fermion mass matrices that
emerges in realistic superstring derived standard--like models.
These models are constructed in the free fermionic formulation
and correspond to superstring models which are based on $Z_2\times Z_2$
orbifold compactification. Among the realistic
 free fermionic models the standard--like
models possess a unique property. They have three and only three chiral
generations. There are no additional generations and mirror generations
that become massive at a large scale. Therefore, the identification
of the light generations is unambiguous. The light generations come
from three distinct sectors, which correspond to the three distinct
twisted sectors of the corresponding orbifold model. The light generations
carry, generational dependent, $U(1)$ charges and Ising model operators.
These symmetries restrict the allowed F--terms in the superpotential, and
are responsible for creating the generations mass hierarchy.

Requiring space--time supersymmetry at the Planck scale restricts the
possible standard--like models. The only models that were found to admit a
supersymmetric vacuum at the Planck scale, are models that allow trilevel
mass terms only for $+{2\over3}$ charged quarks. Similarly,
the requirement of space--time supersymmetry gives a generic choice of
vanishing VEVs, and forces the
Higgs doublets of the third generation to be superheavy.
Consequently, the mass terms for the third generation are suppressed.
This result, like the result for trilevel Yukawa coupling,
is a general characteristic of these models. Therefore, these models
give an unambiguous explanation for the lightness of the lightest generation
relative to the two heavier generations. The suppression of the mass
terms for the lightest generation
states is independent of the specific choice of flat directions
in the cancellation of the anomalous D--term equation.

The following general texture emerges for the fermion mass matrices
in these models,
$$\eqalignno{M_U&=\left(\matrix{\epsilon,a,b\cr
                    {\tilde a},A,c \cr
                    {\tilde b},{\tilde c},\lambda_t\cr}\right)&(18a)\cr
M_D&=\left(\matrix{\epsilon,d,e\cr
                    {\tilde d},C,f \cr
                    {\tilde e},{\tilde f},D\cr}\right)&(18b)\cr
M_E&=\left(\matrix{\epsilon,i,j\cr
                    {\tilde i},E,k \cr
                    {\tilde j},{\tilde k},F\cr}\right)&(18c)\cr}$$
where $\lambda_t=\sqrt{2}g=O(1)$. The entries in capital letters are
diagonal terms which are suppressed by powers of singlet VEVs. The entries
in small letters represent terms which are suppressed by
$({{\Lambda_5}\over{M}})^2$. The diagonal terms for the lightest
generation are suppressed by $({{\Lambda_{Z^\prime}}\over{M}})^2$.
The traditional GUT relations among quark and lepton masses are broken at
various levels of nonrenormalizable terms. At the cubic level the
$SU(5)$ relations are maintained. In model 1, at the quartic level,
the $SU(5)$ relation $m_b=m_\tau$ is obeyed, while in model 2, at the
quintic level it is obeyed only for specific choices of flat directions.
The mixing terms between the generations, in general, are obtained at
different levels of nonrenormalizable terms [\NRT]. Therefore,
the unsuccessful GUT relations for the lighter generation can be cured in the
context of the superstring models.

The analysis presented in this paper provides a qualitative understanding
of the fermion mass and mixing spectrum. The texture of the fermion mass
matrices, exhibited in Eqs. (17), is expected to be a general
characteristic of the class of superstring standard--like models
under consideration. In particular, it is independent of the
specific choice of singlet VEVs, which are used to cancel the D--term
equation of the anomalous $U(1)$. To make progress on a more
quantitative analysis, we must take several steps. First, the
nontrivial corralators of the nonrenormalizable terms have to be calculated
by using well known conformal field theory techniques. Second, the dynamics
of the hidden $SU(5)$ group has to be examined. The scale
$\Lambda_5>{10^{14}GeV}$ depend on the number of fundamental $SU(5)$
representations that are light bellow the Planck scale. The hidden
$SU(5)$ condensates can then be approximated and the mixing between the
generations can be estimated. Finally, the problem of SUSY breaking in the
context of the standard--like models must be addressed and specific
choices of flat directions have to be made, in a phenomenologically
realistic way. The standard--like models provide a highly constrained
laboratory to study these questions, and to address the
question of the origin of fermion masses and mixing in the context of
superstring theory.

\bigskip
\centerline{\bf Acknowledgments}
I would like to thank Lance Dixon and Yossi Nir for useful discussions.
This work is supported in part by a Feinberg School Fellowship.

\refout

\vfill
\eject

\end

\input tables.tex

\hoffset=1.5truein
\nopagenumbers
\magnification=1000
\font\normalroman=cmr10
\font\style=cmr7
\style

\fontdimen12\fivesy=0pt

\textfont0=\sevenrm
\scriptfont0=\fiverm
\textfont1=\seveni
\scriptfont1=\fivei
\textfont2=\sevensy
\scriptfont2=\fivesy

{\hfill
{\begintable
\  \ \|\ $\psi^\mu$ \ \|\ $\{\chi^{12};\chi^{34};\chi^{56}\}$ \ \|\
$y^3y^6$,  $y^4{\bar y}^4$, $y^5{\bar y}^5$,  ${\bar y}^3{\bar y}^6$
\ \|\ $y^1\omega^5$,  $y^2{\bar y}^2$,  $\omega^6{\bar\omega}^6$,
${\bar y}^1{\bar\omega}^5$
\ \|\ $\omega^2{\omega}^4$,  $\omega^1{\bar\omega}^1$,
$\omega^3{\bar\omega}^3$,  ${\bar\omega}^2{\bar\omega}^4$ \ \|\
${\bar\psi}^1$, ${\bar\psi}^2$, ${\bar\psi}^3$,
${\bar\psi}^4$, ${\bar\psi}^5$, ${\bar\eta}^1$,
${\bar\eta}^2$, ${\bar\eta}^3$ \ \|\
${\bar\phi}^1$, ${\bar\phi}^2$, ${\bar\phi}^3$, ${\bar\phi}^4$,
${\bar\phi}^5$, ${\bar\phi}^6$, ${\bar\phi}^7$, ${\bar\phi}^8$ \crthick
$\alpha$
\|\ 0 \|
$\{0,~0,~0\}$ \|
1, ~~~0, ~~~~0, ~~~~0 \|
0, ~~~0, ~~~~1, ~~~~1 \|
0, ~~~0, ~~~~1, ~~~~1 \|
1, ~~1, ~~1, ~~0, ~~0, ~~0 ,~~0, ~~0 \|
1, ~~1, ~~1, ~~1, ~~0, ~~0, ~~0, ~~0 \nr
$\beta$
\|\ 0 \| $\{0,~0,~0\}$ \|
0, ~~~0, ~~~~1, ~~~~1 \|
1, ~~~0, ~~~~0, ~~~~0 \|
0, ~~~1, ~~~~0, ~~~~1 \|
1, ~~1, ~~1, ~~0, ~~0, ~~0, ~~0, ~~0 \|
1, ~~1, ~~1, ~~1, ~~0, ~~0, ~~0, ~~0 \nr
$\gamma$
\|\ 0 \|
$\{0,~0,~0\}$ \|
0, ~~~1, ~~~~0, ~~~~1 \|\
0, ~~~1, ~~~~0, ~~~~1 \|
1, ~~~0, ~~~~0, ~~~~0 \|
{}~~$1\over2$, ~~$1\over2$, ~~$1\over2$, ~~$1\over2$,
{}~~$1\over2$, ~~$1\over2$, ~~$1\over2$, ~~$1\over2$ \| $1\over2$, ~~0, ~~1,
{}~~1,
{}~~$1\over2$,
{}~~$1\over2$, ~~$1\over2$, ~~0 \endtable}
\hfill}
\smallskip
\parindent=0pt
\hangindent=39pt\hangafter=1
\normalroman

{{\it Table 1.} A three generations $SU(3)\times SU(2)\times U(1)^2$ model.
The choice of generalized GSO coefficients is:
$$c\left(\matrix{b_j\cr
                                    \alpha,\beta,\gamma\cr}\right)=
-c\left(\matrix{\alpha\cr
                                    1\cr}\right)=
c\left(\matrix{\alpha\cr
                                    \beta\cr}\right)=
-c\left(\matrix{\beta\cr
                                    1\cr}\right)=
c\left(\matrix{\gamma\cr
                                    1,\alpha\cr}\right)=
-c\left(\matrix{\gamma\cr
                                    \beta\cr}\right)=
-1,$$ (j=1,2,3), with the others specified by modular invariance and
space--time
supersymmetry. Cubic
level Yukawa couplings are obtained only for $+{2\over3}$ charged quarks. }
\vskip 2.5cm

\vfill
\eject

\input tables.tex

\hoffset=1.5truein
\nopagenumbers
\magnification=1000
\font\normalroman=cmr10
\font\style=cmr7
\style

\fontdimen12\fivesy=0pt

\textfont0=\sevenrm
\scriptfont0=\fiverm
\textfont1=\seveni
\scriptfont1=\fivei
\textfont2=\sevensy
\scriptfont2=\fivesy

{\hfill
{\begintable
\  \ \|\ $\psi^\mu$ \ \|\ $\{\chi^1;\chi^2;\chi^3\}$ \ \|\
$y^3y^6$,  $y^4{\bar y}^4$, $y^5{\bar y}^5$,  ${\bar y}^3{\bar y}^6$
\ \|\ $y^1\omega^5$,  $y^2{\bar y}^2$,  $\omega^6{\bar\omega}^6$,
${\bar y}^1{\bar\omega}^5$
\ \|\ $\omega^2{\omega}^4$,  $\omega^1{\bar\omega}^1$,
$\omega^3{\bar\omega}^3$,  ${\bar\omega}^2{\bar\omega}^4$ \ \|\
${\bar\psi}^1$, ${\bar\psi}^2$, ${\bar\psi}^3$,
${\bar\psi}^4$, ${\bar\psi}^5$, ${\bar\eta}^1$,
${\bar\eta}^2$, ${\bar\eta}^3$ \ \|\
${\bar\phi}^1$, ${\bar\phi}^2$, ${\bar\phi}^3$, ${\bar\phi}^4$,
${\bar\phi}^5$, ${\bar\phi}^6$, ${\bar\phi}^7$, ${\bar\phi}^8$ \crthick
$\alpha$
\|\ 0 \|
$\{0,~0,~0\}$ \|
1, ~~~1, ~~~~1, ~~~~0 \|
1, ~~~1, ~~~~1, ~~~~0 \|
1, ~~~1, ~~~~1, ~~~~0 \|
1, ~~1, ~~1, ~~0, ~~0, ~~0, ~~0, ~~0 \|
1, ~~1, ~~1, ~~1, ~~0, ~~0, ~~0, ~~0 \nr
$\beta$
\|\ 0 \| $\{0,~0,~0\}$ \|
0, ~~~1, ~~~~0, ~~~~1 \|
0, ~~~1, ~~~~0, ~~~~1 \|
1, ~~~0, ~~~~0, ~~~~0 \|
1, ~~1, ~~1, ~~0, ~~0, ~~0, ~~0, ~~0 \|
1, ~~1, ~~1, ~~1, ~~0, ~~0, ~~0, ~~0 \nr
$\gamma$
\|\ 0 \|
$\{0,~0,~0\}$ \|
0, ~~~0, ~~~~1, ~~~~1 \|\
1, ~~~0, ~~~~0, ~~~~0 \|
0, ~~~1, ~~~~0, ~~~~1 \|
 ~~$1\over2$, ~~$1\over2$, ~~$1\over2$, ~~$1\over2$,
{}~~$1\over2$, ~~$1\over2$, ~~$1\over2$, ~~$1\over2$ \| $1\over2$, ~~0, ~~1,
{}~~1,
{}~~$1\over2$,
{}~~$1\over2$, ~~$1\over2$, ~~0 \endtable}
\hfill}
\smallskip
\parindent=0pt
\hangindent=39pt\hangafter=1
\normalroman

{{\it Table 2.} A three generations $SU(3)\times SU(2)\times U(1)^2$ model.
The choice of generalized GSO coeficients is:
$$c\left(\matrix{b_j\cr
                                    \alpha,\beta,\gamma\cr}\right)=
-c\left(\matrix{\alpha\cr
                                    1\cr}\right)=
-c\left(\matrix{\alpha\cr
                                    \beta\cr}\right)=
-c\left(\matrix{\beta\cr
                                    1\cr}\right)=
c\left(\matrix{\gamma\cr
                                    1\cr}\right)=
-c\left(\matrix{\gamma\cr
                                   \alpha,\beta\cr}\right)=
-1$$ (j=1,2,3), with the others specified by modular invariance and space--time
supersymmetry. Cubic
level Yukawa couplings are obtained only for $+{2\over3}$ charged quarks. }
\vskip 2cm

\vfill
\eject

\end